# Fossilization in Geopark Araripe studied through X-ray diffraction, scanning microscopy and thermogravimetric analysis

Ricardo J. C. Lima[1], Paulo T. C. Freire[2*], Zélia S. Macedo[3], José M. Sasaki[2], Antônio A. F. Saraiva[4]

[1]Centro de Ciências Sociais, Saúde e Tecnologia, Universidade Federal do Maranhão, Rua Urbano Santos, s/n, 65900-410 Imperatriz – MA, Brazil,

[2]Departamento de Física, Universidade Federal do Ceará, CP 6030, 60455-760 Fortaleza – CE, Brazil,

[3]Departamento de Física, Universidade Federal de Sergipe, Rod. Marechal Rondon s/n 49100000 Aracaju – SE, Brazil,

[4]Departamento de Ciências Físicas e Biológicas, Universidade Regional do Cariri, Rua Cel. Antônio Luiz, 1161, 63105-000 Crato – CE, Brazil.

[*] Fax: 55.85.33669450; e-mail: tarso@fisica.ufc.br

**Abstract**

The Geopark Araripe, located in Northeastern Brazil, is the first UNESCO Natural Park in the South hemisphere and a world-famous fossil deposit of the Early Cretaceous period (approximately 120 million years). Fossilized fish fauna in Geopark Araripe is found inside of sedimentary rocks in three-dimensional forms. In the present study sedimentary rocks and fossil fish Rhacolepis bucalis have been carefully analysed by means of X-ray powder diffraction, scanning electron microscopy and termogravimetric analysis. Mineralogical composition of the fossil fish was explained in terms of facts occurred at the initial stages of the opening of the South Atlantic and the oceanic hydrothermal phenomena ("black smoker", "white smoker" and warm-water events). The occurrence of organic substance was, for the first time, evaluated in collapsed internal elements (intestinal and muscles) by termogravimetric analysis.

# 1. Introduction

The Geopark Araripe (GeoPark ARARIPE, 2006), located in Northeastern Brazil, is an UNESCO Natural Park with impressive traces of continental development, climatic and environmental changes, early life and significant events on the evolution of organisms documented in the rocks of the Sedimentary Araripe Basin, Romualdo Member – Santana Formation (Fara et al., 2005). These sedimentary rocks are famous for the fishes fossil contained into them, registered for the first time in 1831 (Maisey, 1991).

The fossilization process observed in Geopark Araripe region is a very rare event in nature, requiring specific physical and biochemical conditions after burial of the organism and usually it must be aided by geological phenomena (Martill, 1988). Important geologic events in the formation of Romualdo Member - Santana Formation (the fossil deposits of the Geopark Araripe) was uplifted during a geological activity of the Early Cretaceous age (approximately 120 million years) following the rifting apart of Africa and South America to form the South Atlantic Ocean (Maisey, 1991).

In our previous study (Lima et al., 2007), we presented the composition of the fossilized fish scales of the Rhacolepis bucalis which was analyzed by means of X-ray powder diffraction and Fourier transform infrared spectroscopy (FT-IR). The spectroscopic study has proven that the main substance found in the fossilized fish scales is calcium carbonate, $CaCO_3$, and, in lesser concentration it is also found hydroxiapatite, $Ca_5(PO_4)_3(OH)$, evidencing a calcification process.

In the present study we use physical techniques to study a fossil from Geopark Araripe: all mineralogical composition of the extinct Rhacolepis buccalis fossil is investigated by means of X-ray powder diffraction and scanning electron microscopy, the occurrence of organic substance in collapsed internal elements (intestinal and muscles) is evaluated by termogravimetric analysis.

**2. Experimental**

Excavations in Geopark Araripe, the Romualdo Member - Santana Formation (Fig. 1), were carried out near the town of Crato (S 07°17'; W39°23'), in the state of Ceará, Northeast Brazil. The sedimentary rocks (concretion) where the fossilized fishes were found was coded as N1Q1, and deposited in the Paleontology Laboratory of the Universidade Regional do Cariri (URCA).

The measurements of X-ray diffraction of polycrystals were performed through a Rigaku DMAXB model apparatus, which uses a Bragg-Brentano focalization geometry. The interpretation of the diffractogram was performed qualitatively through a so-called 'identification' process. In this stage we employed the computing program Philips X'Pert high score (2001) to identify crystalline phases of the samples. The phases identified were catalogued in data bank of the International Centre for Diffraction Data – ICDD.

Details of the crystalline formations were obtained by scanning electron microscopy (Philips - XL30) and thermogravimetric analysis was made using a *TA Instruments (SDT 2960)* system, under heating rate of 10 °C/min and nitrogen flux of 100 ml/min.

**3. Results and Discussion**

The complete longitudinal section of the sedimentary rocks is shown in Figure 2 (A). We can observe the presence of a fine scale lamination, Fig 2 (B). From the analysis of the scales anatomy the Rhacolepis bucalis species was identified (Lima et al., 2007). In the body cavity it is observed the occurrence of several crystals, Fig. 2 (A-III), and collapsed internal elements (intestinal and muscles), Fig. 2 (A - IV). Also, it is possible to observe evidences of ruptures of the body wall, Fig. 2 (A - V), possibly caused by gas production of bacterial action during an early stage of the fossilization process. It is important to point out that these initial stages in corpse mineralization are influenced by factors such as water temperature, oxygen availabitily, and salinity (Martill, 1988; Maisey, 1991).

The X-ray powder diffraction patterns of the materials are shown in Figure 3. It can be observed that the sedimentary rocks (I), the collapsed internal elements (IV), and the fossilized fish scales (II) have the same structural composition, since the respective diffraction patterns are very similar. The crystalline phase of calcium carbonate (ICDD 85-1108) is predominant for these X-ray powder diffraction patterns.

Crystals grown in the body cavity, Fig. 2 (A-III), present two different colors and morphologies: transparent prismatic and white cubic crystals; their X-ray diffraction patterns are shown in Fig. 3 (III-A) and Fig. 3 (III-B), respectively. Details of the crystalline formations are shown in Figure 4 by scanning electron microscopy. The X-ray diffraction identification points to pure calcium carbonate (ICDD 85-1108) and pure barium sulfate, $BaSO_4$ (ICDD 76-0213) to transparent prismatic crystals and white cubical crystals, respectively.

The hydroxyapatite phase (ICDD 86-0740) was observed in this X-ray diffraction analysis in collapsed internal elements (Fig. 5 - IV) and in fossilized fish scales (Fig. 5 - II) by the occurrence of the peak in $32^o$ approximately (arrow in the Fig. 5 – IV, II), referring to the most intense peak of the crystalline phase of hydroxyapatite phase. In the sedimentary rocks (Fig. 5 - I) and the pure calcium carbonate (Fig. 5 – III A) this peak is absent.

Figure 6 presents the thermogravimetric (Tg) curves of sedimentary rock (curve I) and collapsed internal elements (curve IV). A 9 % weight loss can be observed in curve IV for temperatures between 360 °C and 470 °C, corresponding to the vaporization of organic material(Shin et al., 1999). From this result, it can be concluded that the collapsed internal elements preserve some residual organic material. This is a very interesting result – the occurrence of organic material in the fossils – because the preservation of organic materials occurs generally in ambar, not in sediments (Penney, 2006; Waggoner, 1994). Additionally, a 40 % weight loss is observed between 650 and 810 °C for both samples, corresponding to the degradation of the mineral phases $Ca(CO_3)$ (Koga et al., 1998) and $Ca_5(PO_4)_3(OH)$ (Barinov et al., 2006).

To explain the fossilization observed in the Geopark Araripe region our hypothesis is that an increase in dissolved $CaCO_3$ in the bottom water due to the oceanic hydrothermal phenomena ("black smoker", "white smoker" and warm-water vents) can be associated with catastrophic movement of the opening of the South Atlantic. These phenomena produce precipitation of minerals such as calcium carbonate (Bearman, 1989). Therefore, the fossilization is the result of early infiltration and permeation of tissues by mineral-charged water, where the organic material is subsequently replaced by minerals. Organic material of the fish scales and collapsed internal elements is replaced by $CaCO_3$ (calcification processes). The barite ($BaSO_4$) is also produced in oceanic hydrothermal phenomenon but not participate in the fossilization process. Both, $CaCO_3$ and $BaSO_4$, penetrate in the corpse during the gas exit when the body wall is ruptured. In addition, decomposition of proteinaceous tissues within the unburied fish corpses raised local amine and pH levels to the point where bacteria–related deposition of microcrystalline calcium carbonate could occur.

Phosphate is relatively rare in sea water and generally is not produced by oceanic hydrothermal phenomena. The origin of the hydroxiapatite in fossilized scales and collapsed internal elements is probably biological, as can be realized by the high phosphate levels expected in natural fish bones and scales (Lagler et al., 1962).

**4. Conclusions**

The fossilization of the Rhacolepis bucalis fossils fishes in Geopark Araripe, the Romualdo Member - Santana Formation, is a very exceptional event, which required very special water chemistry, and a remarkable coincidence of events: (i) The oceanic hydrothermal phenomena associated with catastrophic movement of the opening of the South Atlantic producing high levels of dissolved calcium and carbonate ions in water (ii) The proteinaceus tissues (scales and internal elements) forming the body of *Rhacolepis bucalis* became mineralized predominantly by calcium carbonate, calcification process, leading to replacement by chalky, microcrystalline calcium carbonate.

The calcification fossilization process is not complete. Traces of organic material are present in internal elements within the body cavity. This organic material can supply important information and can be used to investigate micro- as well as macro-evolutionary processes, particularly, in molecular level through the investigation of the DNA residues.

Acknowledgments:

We would like to thank CNPq and FUNCAP for financial support. We also acknowledge Dr. J.R. Gonçalves for a reading of the manuscript.

**Caption for figures:**

Fig. 1. Location of the Geopark Araripe (Romualdo Member - Santana Formation).

Fig. 2. (A) Longitudinal section of the sedimentary rock with the body cavity of the Rhacolepis bucalis inside; for numbers I to V, see text. (B) Scales anatomy of the fossil.

Fig. 3. X-ray diffraction patterns of the materials presented in Fig. 2.

Fig. 4. Crystals grown in the body cavity of the fossil, Fig. 2 (A-III).

Fig. 5. X-ray diffraction patterns of the materials in the region 31 – 34 degrees.

Fig. 6. Thermogravimetric curves of sedimentary rock (curve I) and collapsed internal elements (curve IV).

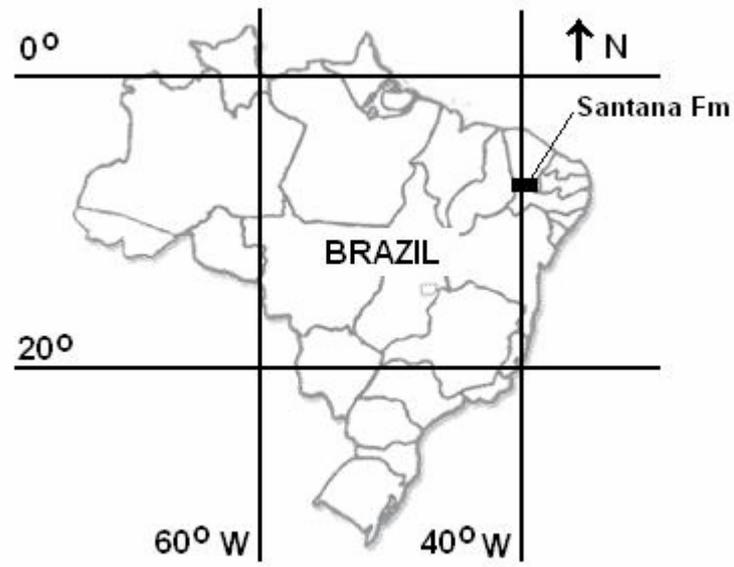

Fig. 1

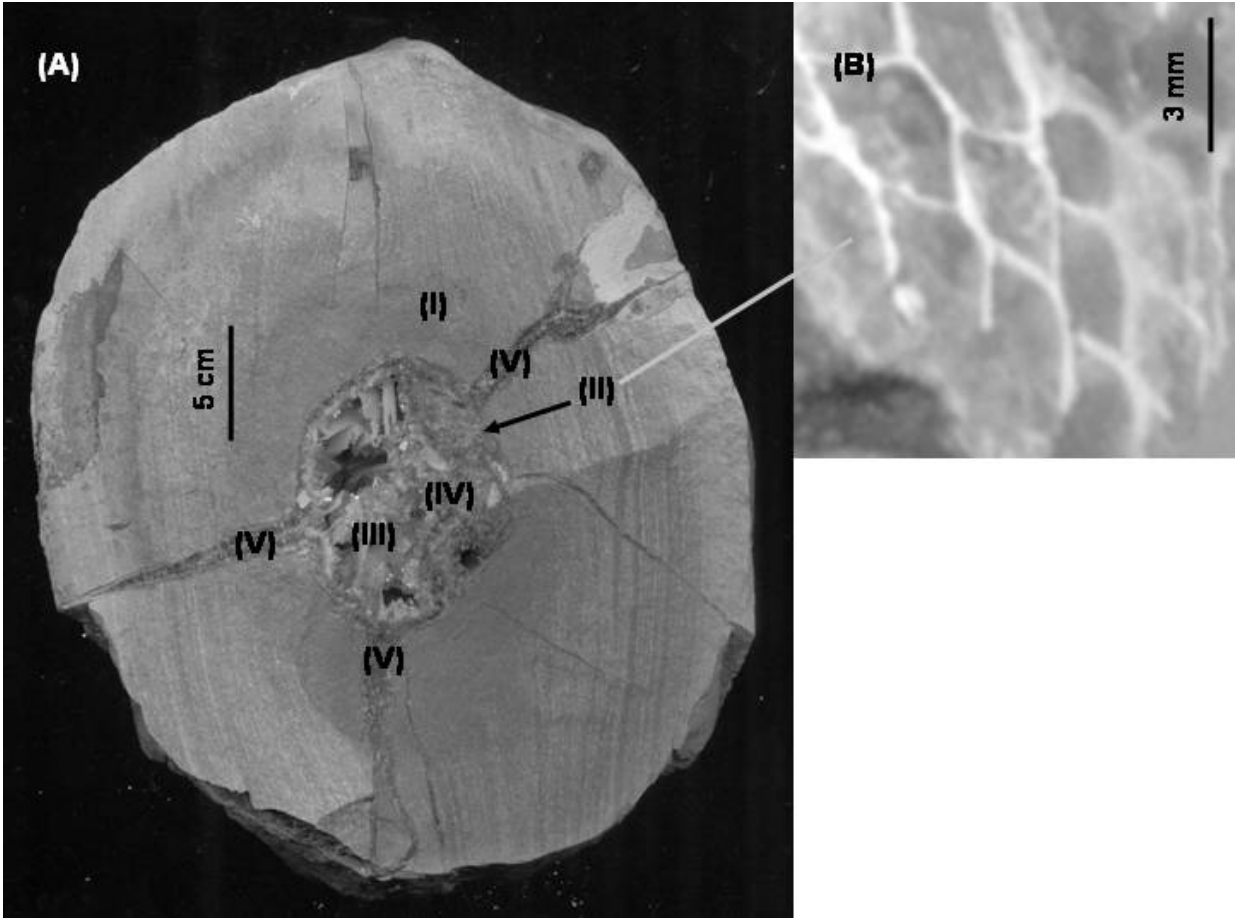

Fig. 2

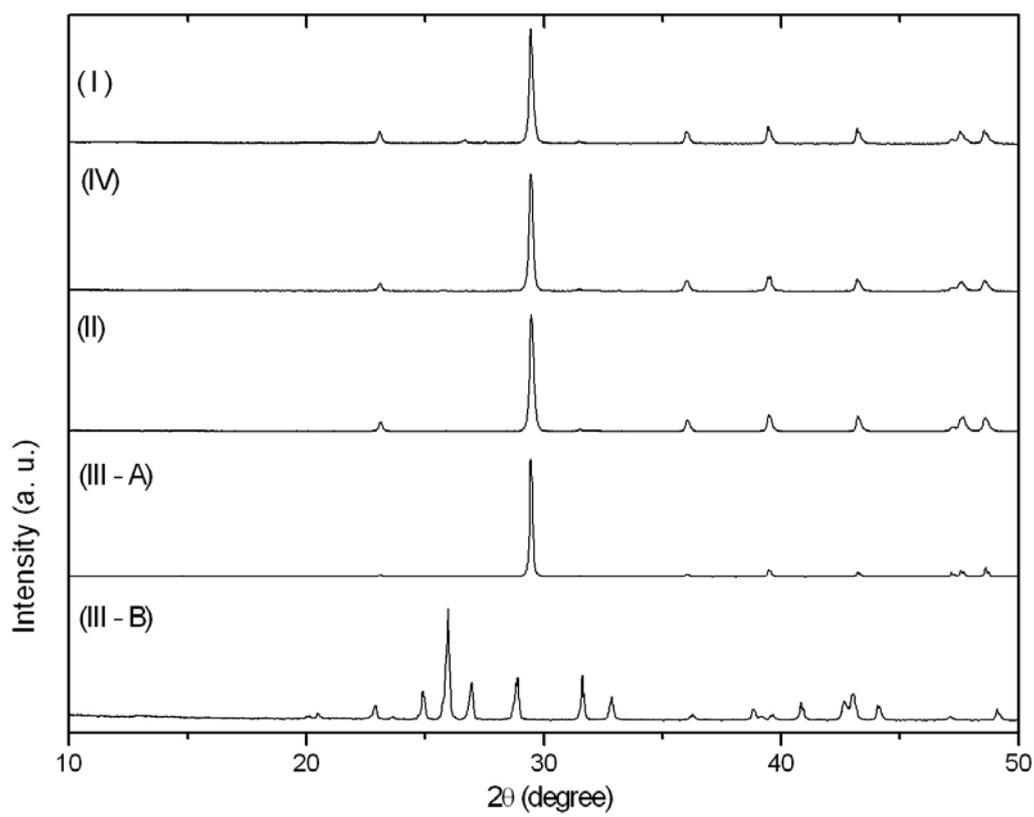

Fig. 3

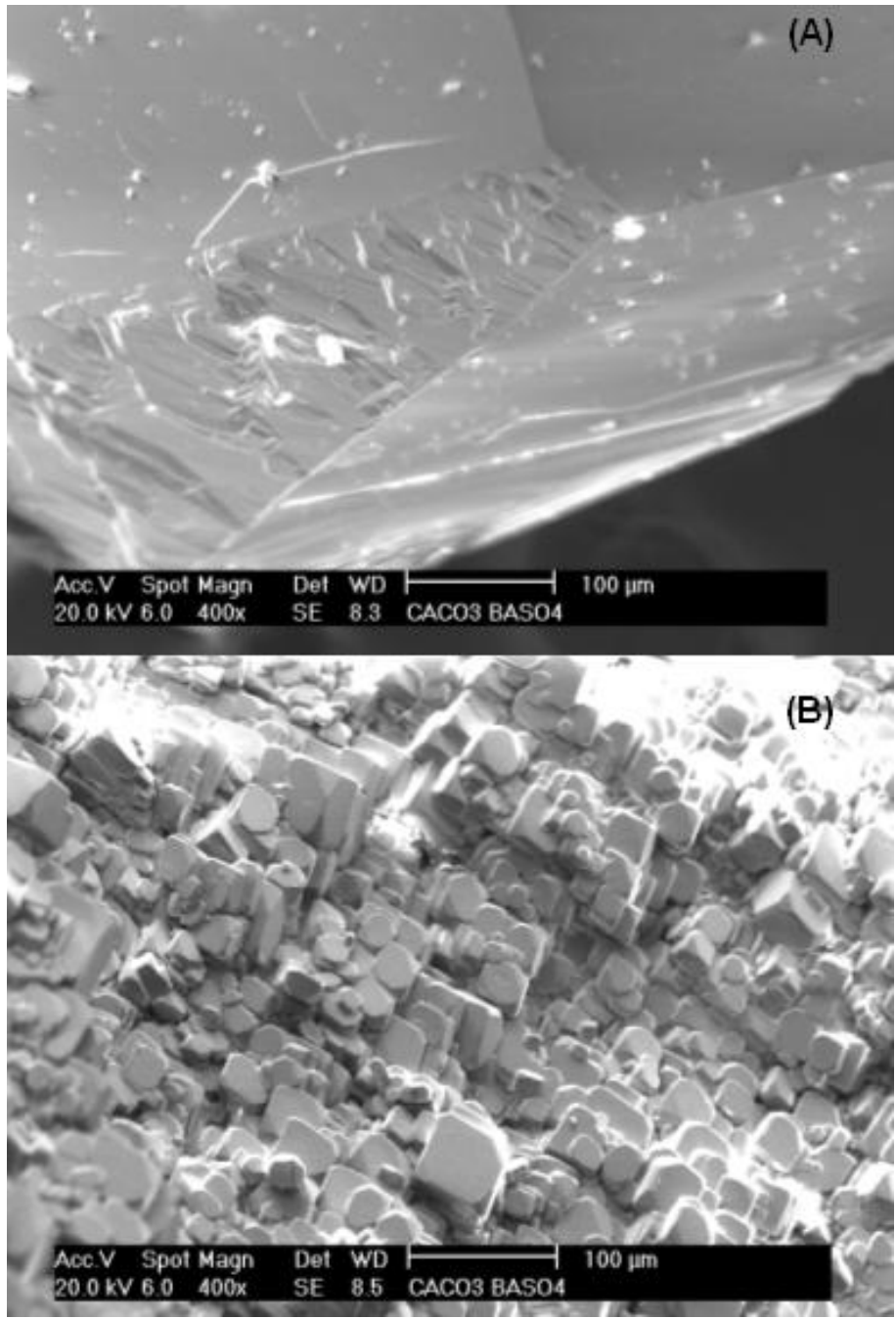

Fig. 4

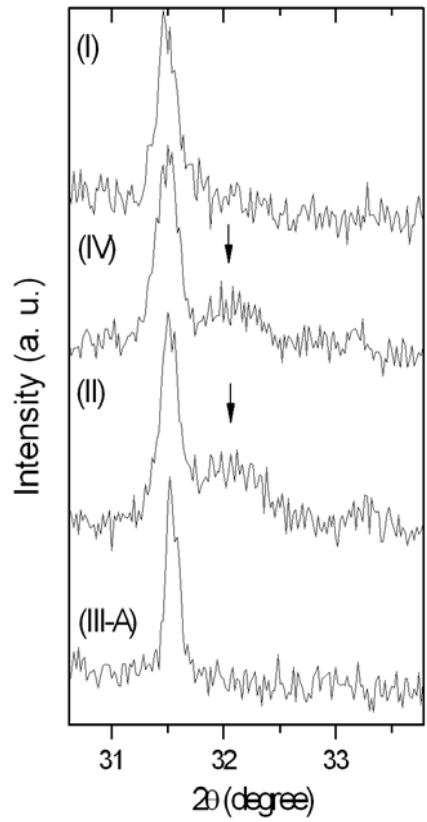

Fig. 5

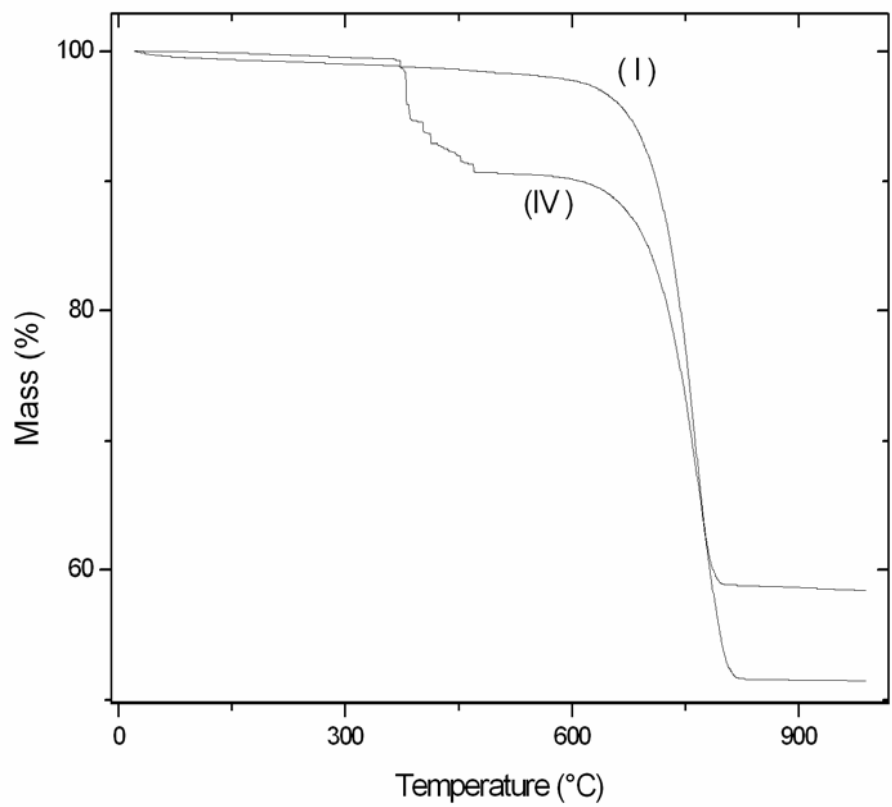

Fig. 6